\documentclass[pdflatex,sn-nature]{sn-jnl}

\usepackage{graphicx}%
\usepackage{multirow}%
\usepackage{amsmath,amssymb,amsfonts}%
\usepackage{amsthm}%
\usepackage{mathrsfs}%
\usepackage[title]{appendix}%
\usepackage{xcolor}%
\usepackage{textcomp}%
\usepackage{manyfoot}%
\usepackage{booktabs}%
\usepackage{algorithm}%
\usepackage{algorithmicx}%
\usepackage{algpseudocode}%
\usepackage{listings}%
\usepackage{pdflscape}
\usepackage{afterpage}
\usepackage{comment}

\lstdefinelanguage{ADQL}{
  morekeywords={SELECT,DISTINCT,FROM,WHERE,AND,BETWEEN,IS,NOT,NULL}, 
  sensitive=true,
  morecomment=[l]{--},    
  morestring=[b]'         
}

\newcommand{\araa}{Annu. Rev. Astron. Astrophys.}   
\newcommand{\aj}{Astron. J.}   
\newcommand{\apj}{Astrophys. J.}   
\newcommand{\apjl}{Astrophys. J. Lett.}   
\newcommand{\apjs}{Astrophys. J. Suppl. Ser.}   
\newcommand{\aap}{Astron. Astrophys.}   
\newcommand{\aapr}{Astron. Astrophys. Rev.}   
\newcommand{\icarus}{Icarus}   
\newcommand{\mnras}{Mon. Not. R. Astron. Soc.}   
\newcommand{\nat}{Nature} 
\newcommand{\pasa}{Publ. Astron. Soc. Aust.}   
\newcommand{\pasp}{Publ. Astron. Soc. Pac.}   

\newcommand{\sub}[1]{_\mathrm{#1}}
\newcommand{\imut}{\iota\sub{mut}}

\theoremstyle{thmstyleone}%
\theoremstyle{thmstyletwo}%

\theoremstyle{thmstylethree}%

\begin{document}

\title{Substantial hot Jupiter formation through galactic perturbations of old wide binary stars
}


\author[1,2]{\fnm{Evgeni} \sur{Grishin}}\email{evgeni.grishin@monash.edu}

\author[1]{\fnm{Jet} \sur{Winter}}

\author[3,4]{\fnm{Jaime A.} \sur{Alvarado-Montes}}

\affil[1]{\orgdiv{School of Physics and Astronomy}, \orgname{Monash University}, \orgaddress{\city{Clayton}, \state{VIC} \postcode{3800}, \country{Australia}}}

\affil[2]{\orgname{OzGrav: Australian Research Council Centre of Excellence for Gravitational Wave Discovery},  \orgaddress{\city{Clayton}, \state{VIC} \postcode{3800}, \country{Australia}}}

\affil[3]{\orgdiv{Australian Astronomical Optics}, \orgname{Macquarie University}, \orgaddress{\city{Sydney}, \state{NSW} \postcode{2109}, \country{Australia}}}

\affil[4]{\orgname{Astrophysics and Space Technologies Research Centre, Macquarie University},  \orgaddress{\city{Sydney}, \state{NSW} \postcode{2109}, \country{Australia}}}

\abstract{The origin of hot Jupiters (HJ), giant planets with orbital period less than $10$ days, is still debated. Recent observations indicate that a significant fraction of them migrated inwards from their formation region late, via long-term high-eccentricity dynamical excitation followed by circularization induced by tidal dissipation. However, current high-eccentricity migration models can only account for $15-30\%$ of the HJ population, overestimating the cold/hot Jupiter ratio by a factor of $\sim2-3$, and cannot explain the inferred high ratio of HJs formed after $2$ Gyr ($40\pm15\%)$. Here we use a dynamical model to show that wide binaries, with semi-major axis $\ge10^3\ \rm AU$, perturbed by the galactic tidal field produce $1.8\pm0.14$ more HJs compared to isolated binary systems, accounting for $26–40\%$ of the total observed population under conservative assumptions. In addition, this wide-binaries pathway is predominantly active in billion-year-old systems, helping to bridge the difference between modelled and observed HJ host-age distribution. We also find that $\sim20\%$ of the systems eject giant planets entirely, resolving the cold/hot Jupiter ratio discrepancy while seeding naturally the population of free-floating giant planets. Our results suggest that wide binaries emerge as important dynamical agents that reshape planetary demographics across billions of years. This hypothesis will be tested by forthcoming exoplanet and microlensing surveys.}

\maketitle
Since the discovery of 51 Pegasi b \cite{51peg}, hundreds of hotJupiters (HJ) have been identified. These giant planets with masses $\gtrsim 0.3\,M_J$ and orbital periods under 10 days form an almost complete magnitude-limited sample around FGK stars \cite{yee2025}, with an occurrence rate of $0.5$–$1\%$ \cite{j10}. HJs are thought to form through three primary channels (see ref. \citep{Dawson_2018} for a review). In the in situ formation scenario, HJs assemble close to their present-day orbits, requiring rapid core accretion and gas envelope growth within $\lesssim 0.1$ AU \citep{batygin2016}. Although this pathway is generally disfavoured due to the limited solid material in the inner disc, it may operate in a subset of systems with tightly packed inner planets, such as WASP-47 \citep{becker2015}. In the disc migration scenario, giant planets form beyond the snow line at a few AU and migrate inward through interactions with the protoplanetary disc, naturally producing early-forming, circular, and spin–orbit aligned systems \citep{lin1996,baruteau2014}. Finally, in the high-eccentricity tidal migration scenario, planets are excited to extreme eccentricities via dynamical processes such as planet–planet scattering or von Zeipel–Lidov–Kozai (ZLK) oscillations induced by stellar or planetary companions, and subsequently circularise through tidal dissipation at pericentre \citep{rasio1996, Wu&Murray}. The occurrence rates and properties of HJs are still puzzling today as no consensus has been reached on the dominant channel of their formation.

Companions to giant planets could shed additional light on the formation channels, where the existence of planetary companions suggest less violent formation pathways. Although $70\pm16\%$ of warm Jupiters (WJ; periods $10-100$ days) have planetary companions, only $12\pm6\%$ of HJs are accompanied by another planet \cite{wu23}. This suggests that the observed population of eccentric WJs is formed via a different channel to HJs, which require to be excited to large eccentricities before they are circularised by tides.

HJs are also found around stars with diverse ages. Earlier detections have found on average HJs in kinematically cold, younger stars compared to the field stars \citep{hamer19, mustill22}, which was used to constrain the efficiency of tidal decay. On the other hand, several HJs have very short tidal decay time which indicates late arrival. In particular, the estimated disruption time of NGTS-10 \cite{ngts-10} of  $\lesssim 40\,\rm Myr$ \citep{Alvarado-Montes2021} is much shorter than its stellar age of $t_{\rm age}=10.4 \pm 2.5\,\rm Gyr$. Similar numbers for WASP 12b also indicate its late arrival \cite{wasp1}. On a population level, Chen et al. (2023) \cite{hj_gyr} found that while HJ occurrence declines modestly with age, a substantial fraction persist around stars older than $\gtrsim 5\,\rm Gyr$. Recent analyses found that about $40\pm15\%$ of HJs form after $2$ Gyr \cite{chen25}, and galactic kinematics suggest that $40–70\%$ of HJs formed after 1.5 Gyr via late-time high-eccentricity migration \cite{old_hj26}.

 Measurements of over $200$ obliquity angles between the stellar spin and the HJ orbital plane via the Rossiter–McLaughlin (RM) effect \cite{obl2024} reveal that hot host stars
($T_{\rm eff} \gtrsim 6300\,\rm K$) exhibit random obliquities, while cool exoplanet hosts show predominantly aligned orbits \cite{obl2022_review}. This pattern supports high-eccentricity migration over disc migration or in situ formation, with subsequent tidal alignment and/or resonant locking \cite{zanazzi25} occurring efficiently in cool stars with convective envelopes, but inefficiently in hot stars with
radiative envelopes. Moreover, the eccentricity distribution of Cold Jupiters aligns with expectations from high-eccentricity tidal migration \cite{weldon2025}. In this scenario, the proto-HJ’s eccentricity is first excited—either by planet–planet scattering \cite{chatterjee2008} or ZLK oscillations \cite{naoz16} driven by stellar companions \cite{Wu&Murray}—and then tidally circularised at pericentre.

Current high-e migration population studies models struggle to explain the observed data. The occurrence rates account for at most $15$–$30\%$ of the observed HJ population \cite{naoz12, petrovich2015, anderson2016}. The ZLK mechanism requires a narrow range of mutual inclinations,
$\imut$, between the planetary and the stellar orbital planes, leaving behind a large population of “Cold Jupiters" ($a > 1\,\rm AU$), with predicted cold/hot Jupiter ratios $\gtrsim 30$, well above the observed ratios of $10$–$15$ \cite{kunimoto2020}. A further challenge is the presence of HJs around old stars,  which suggest late assembly. Chen et al. (2023, 2025) \cite{hj_gyr, chen25} provided a fit for the age distribution of HJ host stars suggest that $\sim 40\%$ of HJ formed after 2 Gyr (which could be as high as $70\%$, \cite{old_hj26}). This age distribution is incompatible with standard ZLK tidal migration (and other models such as secular chaos, as described below).

Recent astrometric data from Gaia has revealed millions of binaries \cite{sample25} and $\sim 10^4$ hierarchical triples \cite{shariat2025}. The recent discovery of a third star in the HAT-P-7 system is a clear evidence of HJ formation in stellar triples with complex chaotic dynamics \cite{yang25}. HJ formation efficiency in higher multiplicity systems is only marginally enhanced \cite{hamers17hj}, with formation timescales much shorter than $\le 1\ \rm Gyr$. Here we demonstrate that the effects of galactic tides (GT) on wide binaries ($a \gtrsim 10^3$ AU) raise HJ occurrence rates by a factor of $1.8\pm0.14$ to $\sim 40\%$ of the observed rate, resolving the cold/hot Jupiter ratio tension and reproducing the observed host age distribution for $t \gtrsim 2.5\ \rm Gyr$. This makes wide-binary interactions a leading channel for forming HJ around old stars. 

\begin{figure*}
    \centering
    \includegraphics[width=0.9\linewidth]{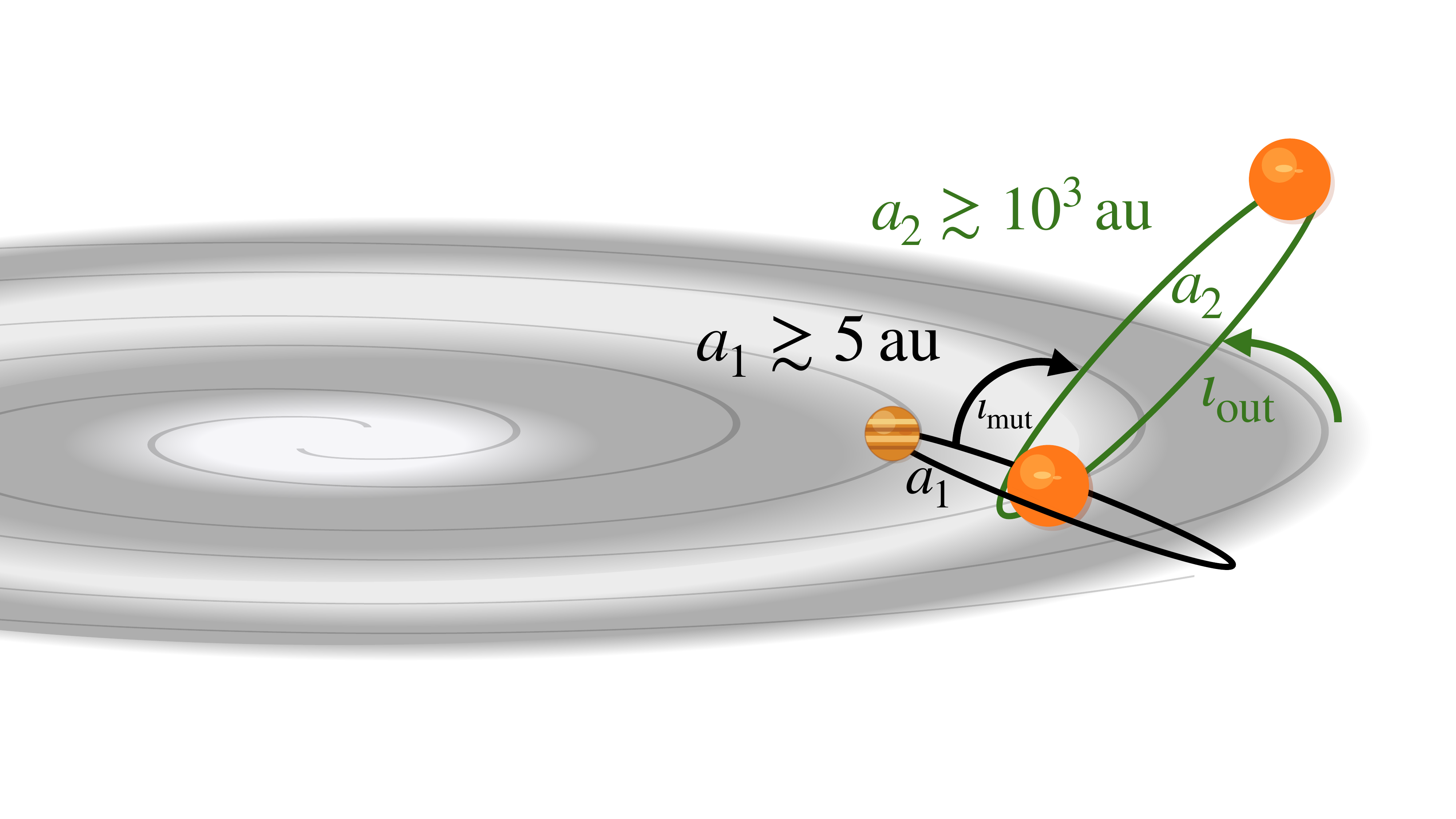}
    \caption{Hierarchical wide three-body system perturbed by the galactic tide. The binary orbit is wide, typically above $a_2 \ge 10^3\ \rm AU$ (green ellipse), and inclined with the galactic plane by an angle $\iota_{\rm out}$. An inner binary giant planet (black ellipse) typically has initial semi-major axis $a_1 \gtrsim 5\ \rm AU$. The mutual inclination between the orbits is $\imut$.}
    \label{fig:1}
\end{figure*}

We set up a hierarchical triple consisting of an inner star-planet binary orbited by a wide stellar companion of separation $a_2 > 10^3$ AU. The wide binary orbit is inclined with an angle $\iota_{\rm out}$ to the galactic plane and is  perturbed by the galactic tide, as sketched in Figure \ref{fig:1}. The mutual inclination between the two orbits is $\imut$. GT can make outer orbits undergo eccentricity oscillations similar to ZLK oscillations, originally applied for studying Sun-grazing Oort-cloud comets \cite{ht86}. For a binary with separation $a_2$ and masses $m_1$ and $m_2$, the timescale of such oscillations is
\begin{equation}
t_\mathrm{GT} \approx  0.89 \left( \frac{m_1+m_2}{M_\odot} \right)^{1/2} \left( \frac{a}{10^4\, \mathrm{au}} \right)^{-3/2} {\rm Gyr}.
\label{Equation: t_galactic tide}
\end{equation}
The maximum eccentricity for the wide binary is $e_{\rm 2,max} = \left(1 - (5/4)\cos^2\iota_{\rm out}\right)^{1/2}$ \cite{ht86}. This can affect the timescale of the inner ZLK oscillations. The qualitative evolution is characterized by the ratio of the secular timescale to the GT timescale \cite{gp22}
\begin{equation} \label{r0}
    \mathcal{R}_0= \frac{t_{\rm sec}}{t_{\rm GT}}
 \approx 1 \left(\frac{M_\odot}{m_2} \right)^{-1} \sqrt{\frac{2m_1}{m_1+m_2}}\left( \frac{a_1}{25\ \rm AU}\right)^{-3/2} \left( \frac{a_2}{10^4\ \rm AU} \right)^{9/2}(1-e_2^2)^{3/2}.
\end{equation}
When $\mathcal{R}_0$ is of order unity, overlapping secular resonances lead the system to evolve chaotically, allowing the planetary orbit to attain extreme eccentricities \cite{gri18b}.

\begin{figure*}
    \centering
    \includegraphics[width=0.99\linewidth]{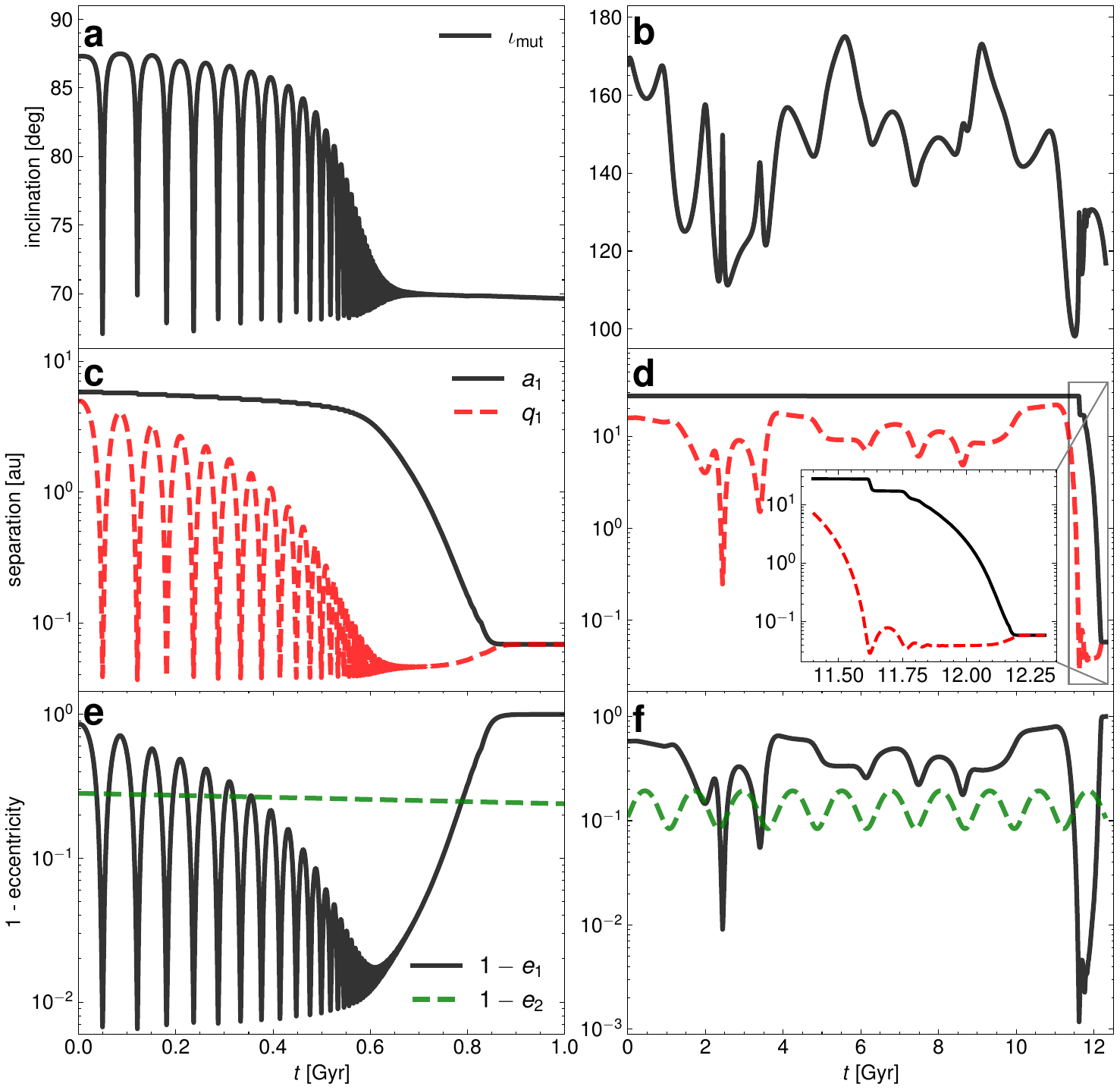}
    \caption{Time evolution for regular and chaotic HJ formation pathways. Panels a., c. and e. on the left show the evolution of the mutual inclination, semi-major axis and pericentre, and the eccentricity of the regular orbit, respectively. Panels b. d. and f. on the right show the same, but for a chaotic orbit. The initial conditions for the regular orbit are \( a_1 = 5.82\,\mathrm{au} \), \( a_2 = 1522.07\,\mathrm{au} \), \( e_1 = 0.15 \), \( e_2 = 0.72 \), \( i_1 \approx 45.9^\circ \), \( i_2 \approx 103.2^\circ \),   \( \iota_{\rm mut} \approx 87.3^\circ \), \( \omega_1 \approx 59.7^\circ \), \( \omega_2 \approx 118.6^\circ \), \( \Omega_1 \approx 45.7^\circ \), \( \Omega_2 \approx 118.5^\circ \), \( m_1 = 1.15\,M_\odot \), and \( m_3 = 0.50\,M_\odot \). The initial conditions for the chaotic orbit are \( a_1 = 27.40\,\mathrm{au} \), \( a_2 = 14540.34\,\mathrm{au} \), \( e_1 = 0.42 \), \( e_2 = 0.89 \); inclinations \( i_1 \approx 138.3^\circ \), \( i_2 \approx 30.2^\circ \), \( \iota_{\rm mut} \approx 167.9^\circ \), \( \omega_1 \approx 261.7^\circ \), \( \omega_2 \approx 192.4^\circ \), \( \Omega_1 \approx 18.7^\circ \), \( \Omega_2 \approx 337.7^\circ \), \( m_1 = 0.66\,M_\odot \), \( m_3 = 0.65\,M_\odot \).}
    \label{fig:2}
\end{figure*}

We evolve the wide triple system using the in-house secular dynamics code \texttt{SecuLab} \cite{gp22}, which solves the secular equations of motion and includes post-Newtonian and tidal effects (see Methods). Figure \ref{fig:2} shows examples of regular and chaotic orbits. Regular orbits must start at small $\cos \iota_{\rm mut}$ to develop sufficiently large eccentricities and form a HJ (a). Tidal dissipation is gradual and occurs at a nearly fixed pericentre $q_1=a_1(1-e_1)$, with several ZLK cycles required to reduce the semi-major axis, consistent with the standard ZLK tidal migration channel \cite{Fabrycky_2007} (c). By contrast, chaotic orbits can begin at any mutual inclination and evolve erratically (b). Most tidal dissipation occurs once the pericentre crosses a critical threshold (d), while $e_1$ undergoes periodic oscillations until tidal decoupling. The outer eccentricity $e_2$ remains nearly constant for regular orbits (e), but undergoes several periodic oscillations in chaotic orbits, driving erratic variations in $e_1$. Efficient tidal dissipation occurs when $e_1$ reaches extreme values ($e_1\approx0.999$). The two channels also differ in their formation timescales: regular orbits form HJs within $\sim1$ Gyr, whereas chaotic orbits evolve over $\sim10$ Gyr. The octupole term can also induce chaotic evolution (see Extended Data Figure \ref{fig:ex1}), but does not reproduce the late arrival times.

\subsubsection*{Analysis and Results}

\textbf{Set-up and results of the population study.} We set up a Monte-Carlo simulation with $10^4$ initial conditions, split evenly between cases with GT turned on and off. Stellar masses are sampled from the Kroupa mass function between $0.5-1.5\ M_\odot$, whereas the planet mass and radius are fixed at $1 \ M_{\rm J}$ and $1 R_{J}$. Cold Jupiters tend to have a small spread in radii around $R_J$ while HJ tend to be more inflated due to tidal heating and intense stellar irradiation (see Extended Data Figure \ref{fig:ex2}). Orbital separations are drawn from log-uniform distributions: $a_1 \sim \log U\ [5, 200]\ \rm AU$ and $a_2 \sim \log U\ [300, 3 \times 10^4]\ \rm AU$, ensuring dynamical stability \cite{ma01}. Note that the lower limit of $300$\ AU goes beyond the wide binary limit of a $1000$ AU, allowing a wider exploration of the parameter space. Mutual inclinations are uniformly distributed in $\cos\imut$. Extended Data Table \ref{table:ICs} lists the full initial conditions. Each system was evolved until reaching either $13.5\ \rm Gyr$ or the main-sequence lifetime of the more massive star, approximated as $10\ (m / M_\odot)^{-2.5}\ \rm Gyr$. In principle, non-adiabatic mass loss \cite{veras14,veras16} and white-dwarf kicks \cite{stephan2024_wd_kick} would enhance HJ formation. The enlarged radius of evolved stars may contribute to planetary engulfment \cite{stephan18,stephan20}. By neglecting these effects and focusing only on main sequence stars, our estimates are conservative lower limits that would only increase for more detailed effects of stellar evolution.

The planet is considered disrupted if its pericentre, $q_1 = a_1 (1 - e_1)$, falls below $2.7\ a_{\rm Roche}$ \cite{gui11}, with the Roche limit calculated as $a_{\rm Roche} = R_J (M_\star / M_J)^{1/3}$. Successful HJ formation at time $t$ is classified by a sufficiently circular orbit, $e_1(t) \leq 0.01$, and a loss of at least $90\%$ of its initial orbital separation, $a_1(t) \leq 0.1 a_1(0)$. Finally, systems can also become dynamically unstable \cite{ma01} as the outer binary evolves due to GT over billion-year timescales.

\begin{table}
\centering
\renewcommand{\arraystretch}{1.7}
\begin{tabular}{|c|c|c|c|c|}
\hline 
\textbf{Galactic Tides} & \textbf{Disrupted} & \textbf{HJs} & \textbf{Warm Jupiters} & \textbf{Unstable}\tabularnewline
\hline 
\hline 
\textbf{On} & $442_{-21}^{+21}$ & $428_{-21}^{+21}$ & $6_{-2}^{+2}$ & $998_{-32}^{+32}$\tabularnewline
\hline 
\textbf{Off} & $388_{-20}^{+20}$ & $239_{-15}^{+15}$ & $7_{-3}^{+3}$ & $0$\tabularnewline
\hline 
\end{tabular}

\caption{Outcome statistics with and without Galactic tides. We classify systems as \textit{disrupted} if they reach low pericentres ($q_1$), as \textit{hotJupiters} if they reach low semi-major axes ($a_1$) and eccentricities $e_1$, \textit{warm Jupiters} if they otherwise classified as stable but have final $a_1< 1\ \rm AU$, and \textit{unstable} if the stability criterion in \cite{ma01} is violated. The $1-\sigma$ errors are from shot noise.}
\label{tab:gt_effects}
\end{table}

 Table \ref{tab:gt_effects} summarises the statistics of each outcome. Both cases yield comparable fractions of $442/5000 = 8.9\%$ and $388/5000=7.8\%$ disruptions, when including and excluding GT, respectively. The GT case produces significantly more successful HJ formation at $(428\pm21)/5000=8.56 \pm 0.42\%$, as compared to $(239\pm15)/5000=4.8 \pm 0.3\%$ when GT are not present. This represents an enhancing factor of $\approx 1.8 \pm 0.14$, indicating that GT predominantly boost secular excitation that drives tidal migration, rather than directly increasing disruption rates. Moreover, the GT case shows that $20 \pm 0.64\%$ of systems become unstable due to changes in the orbit of the wide binary---most likely resulting in planetary ejection that could account for $\sim 20\%$ of known free-floating Jovian planets \cite{sumi23}. In contrast, systems cannot become unstable and form free floating planets without GT, as the outer orbit remains dynamically unperturbed. The number of WJs (systems classified as stable with $a_1<1\ \rm AU$ - which differs from the observational definition of period 10-100 days  \cite{wu23}.) in the simulation is low ($6\pm2$ with GT and $7\pm3$ when GT turned off). This is due to the relatively short duration of the tidal decay (see inset of panel d in Figure \ref{fig:2} where the HJ decays in $\rm 0.5\ \rm Gyr$ and spends less than $0.2\ \rm Gyr$ in the $a_1< 1\ \rm AU$ phase before circularising). This further indicates that WJs and HJs belong to different planet populations.

\begin{figure*}
    \centering
    \includegraphics[width=0.99\linewidth]{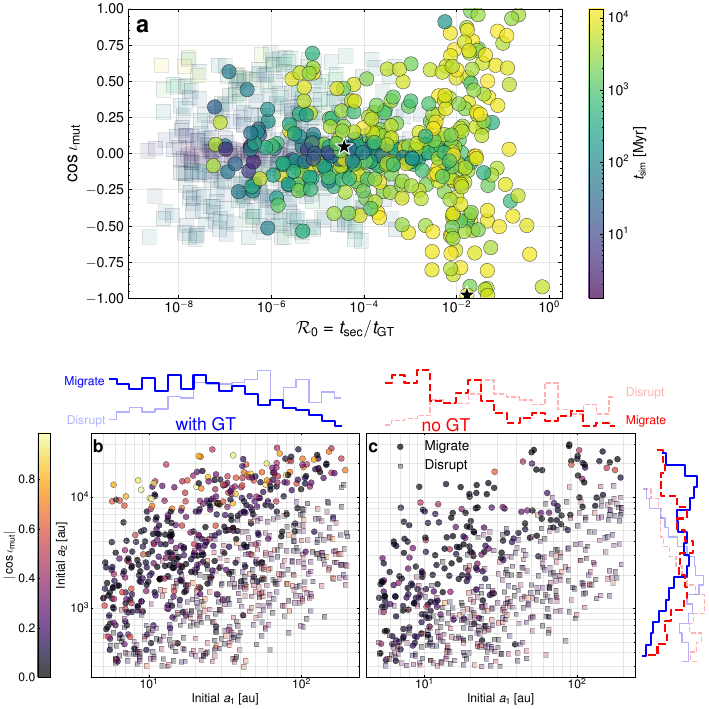}
    \caption{Parameter space of disrupted and migrated systems. Panel a. shows the initial mutual inclination \( \cos \iota_{\rm mut} \) versus $\mathcal{R}_0$ for the run with GT. Coloured circles represent migrated panels and faded squares represent disrupted planets. The colour-code is the total evolution time. Individual simulations in Figure \ref{fig:2} are also shown (stars). Panel b. shows the initial $a_1$-$a_2$ parameter space for the simulations with GT, coloured by the initial mutual inclination between the orbits ($|\cos \iota_{\rm mut}|$). Panel c. shows the same but with GT turned off. Normalised histograms for migrated (thick lines) and disrupted (thin, transluscent lines) systems are overlaid with GT (solid blue, panel b) and without GT (dashed red, panel c).
}
    \label{fig:3}
\end{figure*}

Panel (a) of Figure \ref{fig:3} shows the distribution of disrupted (squares) and migrated (circles) orbits in the $\mathcal{R}_0$–$\imut$ parameter space. Systems marked with stars correspond to those used in Figure \ref{fig:2}. The migrated population exhibits a clear dichotomy: for systems with low $\cos \iota_{\rm mut}$, standard ZLK migration operates on relatively short timescales $\lesssim 1\ \mathrm{Gyr}$, yielding extremely small values of $\mathcal{R}_0$. This also includes systems with moderate $\cos \iota_{\rm mut}$ and low $\mathcal{R}_0$, which are driven by the octupole (eccentric) ZLK \cite{naoz12} (see methods). Once $\mathcal{R}_0$ exceeds a certain threshold, the initial mutual inclination becomes unconstrained and long-term chaotic evolution over $\gtrsim 1\ \mathrm{Gyr}$ dominates. In contrast, disrupted orbits cluster at lower $\mathcal{R}_0$ values and occur predominantly at small $\cos \iota_{\rm mut}$.

Panels (b) and (c) of Figure \ref{fig:3} show $a_1$–$a_2$ correlations, coloured by $|\cos \iota_{\rm mut}|$ overlaid with normalised histograms for both separations in each case. In both simulations, disruptions occur mainly at low $a_2$, where the short secular ZLK timescale ($t_{\rm sec} \propto a_2^3$) prevents tides from damping eccentricity before disruption. HJ migration is favoured at larger $a_2$ when $a_2 \lesssim 100 a_1$, while larger ratios tend to the disruption of HJs. In the GT case (b), an excess of HJs appear at $a_2 \gtrsim 10^4\ \rm AU$ and large $|\cos \iota_{\rm mut}|$, also allowing initially coplanar systems to produce HJs. Without GT (c), wide-orbit HJ formation is sparse and $|\cos \iota_{\rm mut}| \lesssim 0.6$ in most cases. These wide-binary HJs account for the difference seen in Table 1. Due to slow secular evolution, wide binaries rarely lead to the disruption of HJs, so the inclusion of GT does not significantly change disruption rates.

\begin{figure*}
    \centering
    \includegraphics[width=0.9\linewidth]{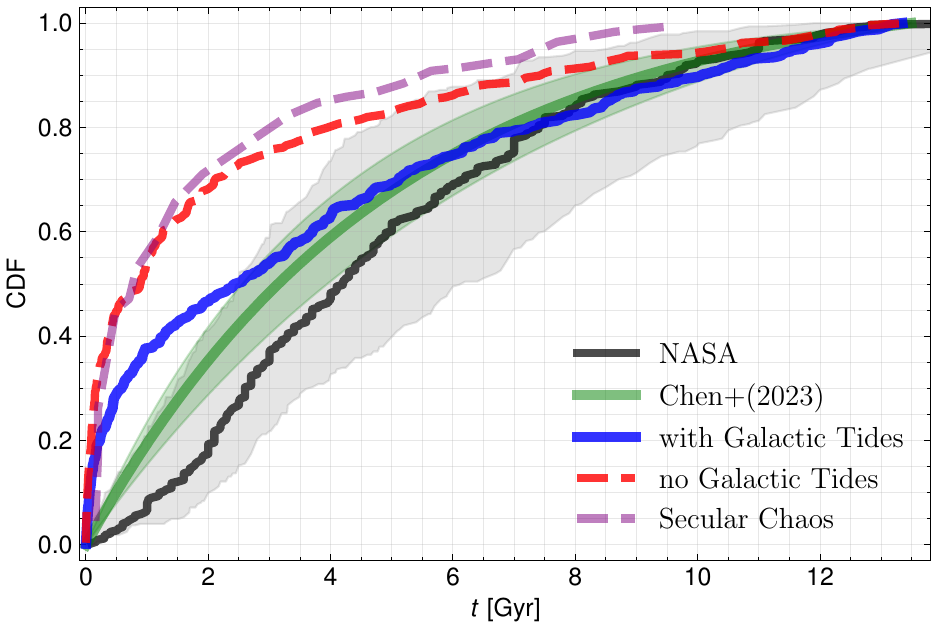}
    \caption{
    Cumulative delay-time distributions of the simulations overlaid with stellar ages and fits. Green line indicates the occurrence rate fit from \cite{hj_gyr}. Black line indicates the HJ sample with known median stellar ages from the \textit{NASA exoplanet archive database}, with 1-$\sigma$ measurement uncertainties represented by the gray shaded area. Dashed purple line shows the result of the secular chaos model \cite{hamers17chaos}. Solid blue line indicates our model with GT while dashed red indicates our model without GT. Figure adapted with permission from ref. \cite{hamers17chaos}, Oxford Univ. Press
}
    \label{fig:4}
\end{figure*}

\textbf{Hot Jupiter formation around old stars.} Figure \ref{fig:4} compares the delay-time distributions of planetary disruptions and HJ formation, with fits to observational data. We see that HJ formation is delayed: only $38\%$ ($55\%$) of HJs form within $1\ \rm Gyr$ with GT on (off). Notably, the GT model produces a significant fraction of very late-forming HJs—$30\%$ occur after $5\ \rm Gyr$—whereas without GT this fraction drops to $17\%$. The secular chaos model was reproduced from Figure 9 in Hamers et al., (2017) \cite{hamers17chaos}, and is comparable to our model without GT.

 To compare with observations, we performed a Table Access Protocol (TAP) search of the public NASA Exoplanet Archive database (\url{https://exoplanetarchive.ipac.caltech.edu/TAP/sync}) for systems with known stellar ages and uncertainties, which returns 319 systems (see Methods). The resulting CDF is shown in black in Figure \ref{fig:4}, with the gray shaded area indicating age uncertainties. For reference, we also plot the bias-corrected fitting formula from \cite{hj_gyr}, where the probability distribution follows an exponential function as $f_{\rm HJ} \propto \exp \left[-0.2 \pm 0.06 (t/\rm Gyr)  \right]$. This corresponds to the cumulative distribution function $\rm CDF(HJ) \propto (1 - \exp\left[-0.2 (t/\rm Gyr)  \right]$. Its CDF and $1-\sigma$ uncertainty are represented by the green line and shaded region, respectively. HJ formation times in simulations with GT are consistent with observed data for $t \gtrsim 2.5\ \rm Gyr$, encompassing $50\%$ of cases. In contrast, without GT only $\sim 6\%$ of HJs fall within the green zone at ages $\gtrsim 9\ \rm Gyr$. The secular chaos model performs worse and does not have any overlap with the green shaded area. Additional analysis excluding the systems with long inspiral times further supports for the wide-binary GT channel as the dominant channel for later arrivals (see Extended Data Figure \ref{fig:ex3}). Therefore,  we conclude that the stellar age reflects the true formation age of HJs and that the wide-binary HJ-formation channel can reproduce most old HJs with stellar ages $\gtrsim2.5\ \rm Gyr$. We note that the oldest systems are around lower mass stars, thus all the data is for main sequence host stars.

\textbf{Estimating the Occurrence rate.} The total occurrence rate can be expressed as $\mathcal{R}_{\rm HJ} = f_{\rm HJ} f_{\rm bin}^{\rm wide} f_{\rm GP}$, where $f_{\rm HJ} \approx 8.6\%$ is the successful HJ formation fraction. The fraction of wide-binary systems $f_{\rm bin}^{\rm wide}$ is the product of the HJ-hosting total binary fraction $f_{\rm bin}$ and the fraction of wide binaries. 
The fraction of companions to HJs is usually taken as $f_{\rm bin}=50\%$ \cite{knutson14, moe2021}, significantly higher than the $20\%$ fraction in the
general solar-type star population \cite{offner23} (although see uncertainties below). The tightest binary systems that host Jovian planets have $a_2 \sim 20\ \rm AU$ \cite{hj_bin}, thus we estimate the wide fraction as $\log_{\rm 10}(3\times10^4/300)/\log_{\rm 10}(3\times10^4/20)\approx 0.63$, yielding $f_{\rm bin}^{\rm wide} \approx 31\%$. $f_{\rm GP} \approx 10{-}15\%$ is the giant planet host fraction. Overall we get $\mathcal{R}_{\rm HJ} \approx 26{-}40\%$ of the observed maximal $\sim 1\%$ occurrence rate. Without GT, this estimate drops to $14-22\%$, which is consistent with previous models (e.g. \cite{petrovich2015, double_HJ}). 

The occurrence rate was derived under conservative assumptions of the high end of the observed rate of $1\%$, which is compatible with the radial velocity (RV) samples. Transiting surveys such as Kepler find smaller occurrence rates of about $0.5\%$. This discrepancy rises due to the exclusion of spectroscopic binaries from the RV sample, while the Kepler sample is not sensitive to binarity. Thus, RV samples indicate the occurrence rate of HJs around single stars while transit surveys indicate occurrence rates around all stars \cite{beleznay2022_occurrence}. Since binarity often precludes planet formation, ref. \cite{moe2021} suggested that the occurrence rates from magnitude-limited surveys underestimate the true occurrence rate by a factor of $\sim 2$. Wide-binaries of $a_2 \gtrsim 10^3\ \rm AU$ are unlikely to hinder planet formation, thus we take the $1\%$ as a fiducial value. Hwang et al. \cite{hwang20} found that planet-host wide binary $>1000 \ \rm AU$ companion fraction has an enhancement of $1.9\pm0.5$ compared to field stars, but this enhancement is weaker if tight binary stars ($<50 \rm\  AU$) are excluded from the field sample. On the other hand, wide binary can have higher than observed companion fractions since 'dual' Jupiters are more likely around both stars due to similar compositions \cite{double_HJ} and due to an unknown sample of wide white-dwarfs or ejected companions \citep{stephan2024_wd_kick}. We thus keep the fiducial $50\%$ companion fractions as a conservative choice without the effects of dual HJs and binary evaporation.

\textbf{cold/hot Jupiter ratio.} The cold/hot Jupiter ratio $y_{\rm wb}$ obtained from the simulations with and without GT is $\approx 7.3 \pm 0.36$ and  $\approx 18.3 \pm 1.18$, respectively. For a rough estimate, the observed cold/hot Jupiter ratio is a sum of the wide-binary and other contributions $y_{\rm obs}= y_{\rm wb}f_{\rm wb} + y_{\rm other} (1-f_{\rm wb})$ where $f_{\rm wb}=0.4$ is the fiducial fraction of HJs formed from the wide binary channel. $y_{\rm other}$ is the unknown cold/hot Jupiter ratio obtained from all other channels. Disc migration can yield the observed cold/hot Jupiter ratios \cite{coleman16}, whereas other channels such as planet-planet scattering and dynamical ZLK-like channels have much less efficient HJ formation yields (or much higher cold/hot Jupiter ratio) \cite{Dawson_2018} For a fiducial $y_{\rm other}=20$ we have $y_{\rm obs}=0.4 y_{\rm wb} + 12$, which is $\lesssim 15$ for the GT model, but closer to $20$ for the model without GT. Although this is only a rough estimate, the GT model matches with the observed ratio of $y_{\rm obs}=10-15$ where the other model without GT is not efficient enough to reduce $y_{\rm obs}$.

\textbf{Free Floating Giant Planets (FFGP).} The rate of free-floating giant planets (FFGP) is  
$20\% \times 0.15 \times 0.5 \approx 1.5\%$.  
The observed FFGP rate per star is estimated by  
$dN/d\log_{10} M_p = 2.18^{+0.52}_{-1.40} \times \left(M_p/8 M_\oplus\right)^{-p}$, where $p = 0.96^{+0.47}_{-0.27}$ \cite{sumi23}. For Jupiter-mass planets, ejection in wide binaries accounts for approximately $24^{+117}_{-15}\%$ of the population. The efficiency of this ejection mechanism is not reduced for larger Jupiter masses, so it is expected to remain robust for masses above Jupiter’s, potentially explaining  
the higher-mass tail of FFGP. Planet-planet scattering  produces mostly free-floating rocky planets \cite{hareesh2025}, however additional processes may need to be considered to account for the galactic FFP population \cite{veras12}.

\textbf{Comparison with Stellar flybys.} Binaries wider that several $1000$'s of AU are prone to perturbations from stellar flybys. In our case, the upper limit of our simulations ($3\cdot 10^4\ \rm AU$) matches the binaries that have a mean lifetime of $10\ \rm Gyrs$, and is further supported by the observed depletion of binaries of projected separations beyond $0.15\ \rm pc$ (see Methods for details). Diffusive processes can also alter the outer orbit to highly eccentric with pericentre $q_2 \ll a_2$. If the pericentre falls below a critical threshold $q_2 \le 3 a_1$, then dynamical instabilities may occur which most likely lead to an ejection. We have calculated the probability for such close encounter from diffusion in the Methods, which yield $\approx 5.5\%$ for our choice of parameters. Coherent eccentricity oscillation from GT that leads to an ejection occurs for $20\%$ of our simulations. Thus, dynamical instabilities are $3-4$ times more likely in the case of GT, which are accounted for. This is in line with the original estimate of ref. \cite{ht86} that estimated that the contribution from the galactic tide will dominate over stellar flybys in producing Sun-grazing Oort cloud comets. 

\textbf{Limitations.} Our study adopts well-established equilibrium tide models and spin pseudo-synchronisation prescriptions \cite{hut81} with fixed apsidal motion constants. We have confirmed the robustness of our results through an additional test run of HD80606b with different dynamical tide prescriptions (Extended Data Figure \ref{fig:ex4}) and additional $5,000$ simulations including dynamical tides for highly eccentric binaries (see Methods). We also do not explicitly evolve the obliquity angles. This is justified as planetary spins rapidly synchronise and do not affect the tidal evolution. The stellar spin can vary chaotically \cite[e.g.,][]{anderson2016, hamers17hj}. We expect that the obliquity distribution will be similar to the '$3+1$' case in Figure 9 of \cite{hamers17hj}, which is more uniform between $30-150$ deg and does not have a strong peak near the ZLK angles (40 or 140 deg) as in the standard ZLK case \cite{Fabrycky_2007}. The period distribution matches only for periods beyond $4.5$ days, while there is an abundance of shorter periods in the simulations (Extended Data Figure \ref{fig:ex5}). This discrepancy can be due to the dominance of lower mass stars in our initial conditions, which differs from the FGK Kepler sample, and due additional tidal effects that destroy HJs on shorter period, which are unaccounted for (see methods). Our galactic tide treatment captures the dominant vertical component. More detailed galactic tidal models and the effects of occasional stellar encounters have been explored for wide binaries by Stegmann et al. (2024) via direct N-body simulations \cite{stegmann24}. However, applying such simulations to our systems---where the inner orbital timescales are much shorter and need to be resolved over $\gtrsim 10\ \rm Gyr$---remains extremely challenging. 

\subsubsection*{Discussion and conclusions}
We studied the combined effects of galactic tides (GT) on wide stellar binaries and von-Zeipel-Lidov-Kozai (ZLK) oscillations and tidal decay on the inner giant planets. Our novel channel forms hotJupiters (HJ) predominantly at later times, successfully reproducing the delay-time distribution of HJs around old stars with $t_{\rm age} \gtrsim 2.5\ \rm Gyr$. We found an enhancement factor for HJ formation of  $1.8\pm0.14$ compared to the model without GT. This, in turn, can explain $26-40\%$ of all HJ and about $\sim 25\%$ of free-floating giant planets (with large uncertainties). This is likely a lower limit since HJs around single stars could have lost the wide binary companion due to stellar evolution \cite{stephan2024_wd_kick}. Our channel is also consistent with the observed cold/hot Jupiter ratios where dynamical instabilities eject a significant fraction of giant planets, hence also contributing to the free-floating giant planet (FFGP) population. Future studies of more detailed tidal evolution, galactic potential, and planetary structure will refine the formation rates and properties. This wide-binary channel can explain not only FFGPs and HJs, but also ultra-short-period planets whose formation rate increases with stellar age \cite{usp_age}.

\renewcommand{\figurename}{Extended Data Fig.}
\setcounter{figure}{0}
\renewcommand{\tablename}{Extended Data Table}
\setcounter{table}{0}

\section*{Methods}

\subsection*{Secular dynamics} 

\textbf{von Zeipel-Lidov-Kozai oscillations} 

Consider an inner binary consisting of bodies $m_1$ and $m_2$ with semi-major axis $a_1$ and eccentricity $e_1$, the outer binary consists of the centre of mass on the inner binary and the third body $m_3$, with $a_2$ and $e_2$. The mutual inclination between the orbital planes is $\imut$. ZLK oscillations are large-amplitude exchanges between the eccentricity and inclination of the inner orbit, driven by the torque of the outer perturber that occur on long, secular time \citep{antognini15}
\begin{equation}
    t_{\rm sec} = \frac{8(1-e_2^2)^{3/2}}{15\pi} \frac{m_{\rm tot}}{m_3} \frac{P_{\rm out}^2}{P_{\rm in}} = \frac{16}{15}\frac{m_{1}^{1/2}}{G^{1/2}m_{3}}\frac{a_{2}^{3}(1-e_{2}^{2})^{3/2}}{a_{1}^{3/2}}
    \label{Equation: t_secular}
\end{equation} 
where $G$ is the gravitational constant, $m_{\rm tot}=m_1+m_2+m_3$ is the total mass and we assume that the planetary mass is small, $m_2 \ll m_1, m_3$.

After double averaging over both orbits and truncating at the leading quadrupole order $\propto(a_1/a_2)^2$ in the test-particle limit, these oscillations preserve the inner orbit's semi-major axis $a_1$ and the z-component of the inner angular momentum $j_z=\sqrt{1-e_1^2}\cos \iota_{\rm mut}$ and the problem is integrable. The maximal eccentricity depends mostly on the mutual inclination (for initial quasi-circular orbit, see review of Naoz (2016) \cite{naoz16}). 
\begin{equation}
    e_{\rm max} = \sqrt{1 - \frac{5}{3}\cos^2 \iota_{\rm mut}}
\end{equation}
However, for weak hierarchies, the double averaging approximation breaks down \citep{luo16,tremaine23,kk24a}, and the expression for $e_{\rm max}$ can be corrected for expansion in the period ratio to first order \citep{gpf18,man22,gri24I,gri24II} and second order \citep{lei25a, lei25b}. 

At octupole order, $j_z$ is no longer conserved and the evolution is chaotic, and orbital flips and extreme eccentricity excitations can occur for a wider range of parameter space \citep{naoz12, naoz13,lml15}. The octupole parameter that governs the dynamics is:
\begin{equation}
    \epsilon_{\rm oct} = \frac{a_1 e_2}{a_2 (1 - e_2^2)},
\end{equation}
and the corresponding octupole timescale is \citep[ref. ][Eq.~79]{antognini15}:
\begin{equation}
    t_{\rm oct} =\frac{256\sqrt{10}}{15\pi\sqrt{\epsilon_{\rm oct}}}t_{\rm sec} \approx 17 \frac{t_{\rm sec}}{ \epsilon_{\rm oct}^{1/2}} \propto \frac{a_{2}^{7/2}(1-e_{2}^{2})^{2}}{a_{1}^{2}e_{2}}.
    \label{t_octopole}
\end{equation}
Although the excitation to high eccentricity occurs more frequently, more planetary disruption is $3-7$ times more likely than HJ formation, depending on initial conditions \citep{petrovich2015,anderson2016}.

\begin{figure*}
    \centering
    \includegraphics[width=0.99\linewidth]{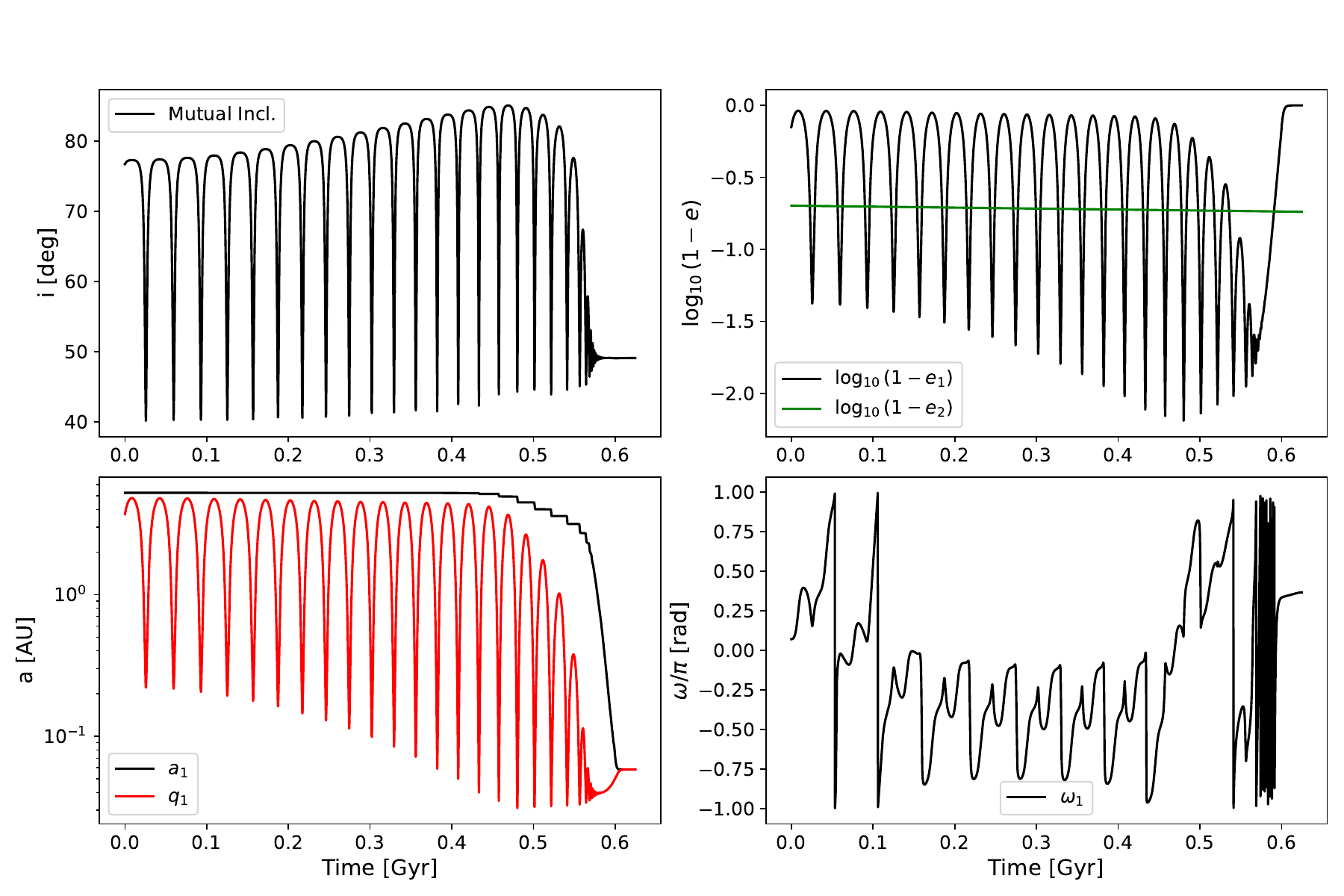}
    \caption{Example of time evolution of a representative system which represents octupole-level ZLK dynamics.  
The initial conditions are: semi-major axes \( a_1 = 5.25\,\mathrm{AU} \), \( a_2 = 1236.74\,\mathrm{AU} \); eccentricities \( e_1 = 0.29 \), \( e_2 = 0.80 \); inclinations \( i_1 \approx 48.1^\circ \), \( i_2 \approx 107.3^\circ \), and mutual inclination \( \iota_{\rm mut} \approx 76.75^\circ \).  
The arguments of pericentre are \( \omega_1 \approx 12.8^\circ \), \( \omega_2 \approx 210.7^\circ \); and the longitudes of ascending node are \( \Omega_1 \approx 117.5^\circ \), \( \Omega_2 \approx 170.5^\circ \).  
The masses are \( m_1 = 0.77\,M_\odot \), \( m_2 = M_J \), and \( m_3 = 0.55\,M_\odot \)}
    \label{fig:ex1}
\end{figure*}

Extended Data Figure \ref{fig:ex1} illustrates the enhanced eccentricity excitations due to the octupole effects. Contrary to the standard (quadrupole) ZLK, The 'Kozai constant' $\sqrt{1-e_1^2}\cos \iota_{\rm mut}$ is no longer conserved and the pericentre of the inner binary is gradually descends, as expected analytically \citep{weldon_naoz2024}. From Eq. 26 in \citep{weldon_naoz2024} we have $t_{\rm descent} \approx \Upsilon \eta t_{\rm oct}$. In our case, $t_{\rm oct} \approx 174 t_{\rm sec}$,  $\eta\approx \sqrt{2r_p / a_1}/7$ and $\Upsilon\approx7$ (see their appendix B), so $t_{\rm descent}\approx0.074t_{\rm oct} \approx 15 t_{\rm sec}$, consistent with the number of ZLK cycles required to reach $0.03\ \rm AU$. The inclination range in this regime is larger since extreme eccentricity can be obtained more gradually. The argument of pericentre also evolves more chaotically with alternating phases of libration and full circulation. The timescale is also longer and takes $\sim 20 t_{\rm sec}$ to get decoupled compared to the much faster evolution in the quadrupole ZLK case.

\textbf{Galactic Tides} 

The tidal field of the galactic disc can perturb the orbits of wide binaries. A sketch of the wide triple system in the galactic disc is depicted in Figure \ref{fig:1} in the main text. To leading order, the vertical tide is given by the potential \cite{ht86, gp22}:
\begin{equation}
U(x, y, z) = - \frac{G m_{\text{tot}}}{\sqrt{x^2 + y^2 + z^2}} + 2 \pi G \rho_0 z^2,
\label{Equation: Galatic potential}
\end{equation}
where $\rho_0 = 0.185\, M_\odot\,\mathrm{pc}^{-3}$ is the local density and $x,y,z$ are Cartesian distances from the galactic centre and $m_{\text{tot}}$ is the total mass of the binary. After orbit-averaging, the corresponding secular Hamiltonian is given by:
\begin{equation}
\mathcal{H}_{\text{GT}} = -\frac{G m_{\text{tot}}}{2a_2} + \pi G \rho_0 a_2^2 \sin^2 \iota_{\rm out} \left( 1 - e_2^2 + 5e_2^2 \sin^2 \omega_2 \right),
\label{Equation: Galactic Hamiltonian}
\end{equation}
where $\iota_{\rm out}$ is the angle between the outer binary and the galactic plane, and $\omega_2$ is the argument of pericentre of the outer orbit. 

The secular evolution and phase space are morphologically similar to the ZLK structure \cite{ham19}; the critical inclination for the onset of the oscillations is $\cos^2 \iota_c < 4/5$, or $\iota_c \approx 26.56^\circ$. For a given prograde inclination $\iota_{\rm out} > \iota_c$ (with the equivalent reflection symmetry for retrograde inclinations), the maximal eccentricity is $e_{\rm max} = (1 - (5/4)\cos^2\iota_{\rm out})^{1/2}$. Developing high eccentricity in wide binaries was first demonstrated by Heisler and Tremaine (1986) \cite{ht86} in the context of Sun-grazing comets emanating from the Oort cloud.

The timescale for secular evolution driven by the galactic tide is given by \citep{ht86, gp22} and provided in Equation \ref{Equation: t_galactic tide} in the main text. The ratio $\mathcal{R}_0$ between the secular timescale and the GT timescale is given by Equation \ref{r0} in the main text. Chaotic evolution and extreme eccentricity is possible when the ZLK and galactic tide timescales are comparable ($\mathcal{R}_0 < \,1$). For further details, discussion and astrophysical implications on the coupled evolution of wide stellar triples with galactic tides we refer to Grishin and Perets (2022) \cite{gp22}. Here we focus on formation of HJs in wide binaries and use similar methods of population synthesis.

\subsection*{Population Synthesis} \label{sec: Population Synthesis}

\textbf{Numerical methods and code implementation} \label{subsec:numerical}

To study the role of chaotic dynamics in HJ formation via high-e migration in wide triple systems, we employ the secular evolution code \texttt{SecuLab}, available at \url{https://github.com/eugeneg88/SecuLab} \cite{gp22}. \texttt{SecuLab} solves the secular equations of motion for hierarchical triple systems up to octupole order, which uses the orbit-averaging principle and thus is faster than direct N-body codes. Additional corrections due to the Brown Hamiltonian \citep{luo16,tremaine23,gri24I} are also implemented. We've implemented Luo et al., (2016)'s version \cite{luo16}, while Tremaine (2023) \cite{tremaine23} shows that the various different forms are equivalent up to short term oscillations. We use the eccentricity and normalised angular momentum vector elements $\boldsymbol{e}$, $\boldsymbol{j}$ \citep[e.g.][]{tremaine09, petrovich2015} for both orbits rather than the traditional Delaunay elements \citep[see details and discussion in ref.][]{lml15}. The external perturbation from the galactic tide changes the invariable plane of the triple system, which causes inconsistencies in the Delaunay variables \citep{naoz13}, while the vector elements are not sensitive to the choice of reference frame.  We choose the reference frame to align with the galactic plane. The equations of motion for the inner orbit are similar to the ones used in Petrovich (2015) \cite{petrovich2015}, but without the spin vectors, which are assumed to be pseudo-synchronised. For the outer orbit, the equations of motion have both feedback from the inner orbit (although it is negligible for the test particle limit) and the effect from the galactic tide Hamiltonian (Equation \ref{Equation: Galactic Hamiltonian}).

Beyond gravitational three-body interactions, \texttt{SecuLab} accounts for perturbations due to the galactic tide on the outer orbit \cite{gp22} and additional conservative and dissipative forces. The conservative forces include 

i) General-Relativistic (GR) precession of the inner binary from the 1st post-Newtonian (PN) term.

ii) Apsidal precession due to tidal bulges on both the planet and star. The extra precession may limit the maximal eccentricity attained during the evolution \citep[see details in][]{lml15}. Mangipudi et al. (2022) \cite{man22} derived analytical expressions for the maximal eccentricity including the 1PN term only, applicable for compact objects, while \cite{vg2025} extended the expression to include tidal bulges applicable for stars.

The dissipative forces include tidal friction \citep{hut81,Fabrycky_2007} and gravitational-wave emission, occurring at the 2.5PN order. In our case for fluid bodies, gravitational-wave emission is negligible compared to tidal dissipation. We use the equilibrium tide model \citep{hut81}. We do not evolve the spin or the planet, but rather assume that the planetary spin is pseudo-synchronised with the orbit, which occurs much faster that tidal circularization \citep{hut81}. We discuss the implementation of dynamical tides and its effects on the results in later sections.

\subsection*{Initial conditions of the population study}\label{sec: initial conditions}

\begin{table} 
\centering
\renewcommand{\arraystretch}{1.2}
\begin{tabular}{@{} l l @{}}
\toprule
\textbf{Parameter} & \textbf{Distribution / Value} \\
\midrule
$m_1$, $m_3$ & Power law: $f(m) \propto m^{-2.3}$, $m \in [0.5, 1.5]\,M_\odot$ \\
$m_2$ & $M_{\mathrm{J}} = 0.0009546\,M_\odot$ \\
$r_1$, $r_3$ & Mass-radius relation \\
$r_2$ & $R_{\mathrm{J}} = 0.10276\,R_\odot$ \\
$a_1$ & $\log U \sim [5, 200]$ AU \\
$a_2$ & $\log U \sim [300, 3 \times 10^4]$ AU \\
$e_1$ & Rayleigh: $f(e)=(e/\sigma^2)\exp(-e^2/2\sigma^2);\quad \sigma = 0.2$ \\
$e_2$ & Thermal: $f(e) = 2e$ \\
$\iota_1$, $\iota_2$ & $\cos \iota \sim U[-1, 1]$ \\
$\omega_1$, $\omega_2$ & $U[0, 2\pi)$ \\
$\Omega_1$, $\Omega_2$ & $U[0, 2\pi)$ \\
$k_{\star}$, $k_p$ & $0.028$, $0.37$ \\
$t_{\nu \star}$, $t_{ \nu,p}$ &  $50\ \rm yr$, $0.01\ \rm yr$ \\
\bottomrule
\end{tabular}
\caption{Initial conditions for the simulations. Masses $m_1$ and $m_3$ are drawn from a power-law distribution with index $-2.3$ in the range $[0.5, 1.5]\,M_\odot$, while $m_2$ is fixed to one Jupiter mass. Radii $r_1$ and $r_3$ are calculated via a mass-radius relation, and $r_2$ is fixed to Jupiter's radius. semi-major axes $a_1$ and $a_2$ are drawn from log-uniform distributions. Eccentricity $e_1$ is sampled from a Rayleigh distribution with $\sigma=0.2$, and $e_2$ from a thermal distribution. Angular parameters are uniformly sampled.}
\label{table:ICs}
\end{table}

In order to examine the role of the galactic tide in forming HJs, we run $N=10^4$ dynamically stable systems according to Mardling and Aarseth (2001)'s stability criterion \cite{ma01}. Mardling and Aarseth (2001)'s \cite{ma01} criterion is applicable for triples of similar masses. Later studies updated the dependence with the mutual inclination and more extreme mass ratios \citep{grishin2017, tory22, vynatheya22}, with initial conditions drawn as per Exended Data Table 1. We run $1/2$ of the simulation with GT and the other $1/2$ without GT (so each run has $5,000$ initial conditions). Extended Data Table 1 lists our choices for the initial distribution of the parameters. We motivate our choices for the population study below:  

\textbf{Masses:} The occurrence rate of HJs is dependent on the survey, its biases and completeness. The Kepler mission targeted primarily FGK stars, which are intrinsically brighter, easier to monitor with high photometric precision, and more amenable to RV follow-up than faint, active M dwarfs. RV surveys tend to find an increase of the occurrence rate with the star. However, recent analysis of TESS found that HJs are more common around lower mass GK stars which peak at around $0.8M_\odot$ \cite{beleznay2022_occurrence, gan24_rates}. The occurrence rates around M dwarfs in the range of $0.45-0.65 M_\odot$ is much lower, about $0.27\pm0.09\%$  \cite{gan23_md}. This trend is, however, most likely determined by the metallicity \cite{gan25_metallicity}, since HJ occur around more metal-rich stars \cite{j10} and more massive stars are more diverse in their metallicity. Having no definitive observed trend with the stellar mass, we remain agnostic to the initial mass distribution and simply sample the stellar masses from The Kroupa mass function ($m_1, m_3 \propto m^{-2.3}$) and  use the mass radius power law $R \propto m^\xi$ with $\xi = 0.8$ for $m<M_\odot$ and $\xi = 0.57$ for $m>M_\odot$ \citep{torres2010}. The planetary masses and radii are kept at the Jovian values. The mass does not play a role in the dynamics since the system is always in the test particle limit. Also, the radius is almost independent of the mass (see more details and discussion in the mass-radius relation section).

\textbf{Semi-major axes:} For the wide stellar binaries the distribution of semi-major axes is consistent with earlier studies \citep{d91} and more recent data available from Gaia \citep{gaia21}. The spatial distribution of giant planets, especially beyond the snow ice at a few astronomical units, is less constrained. According to \cite{fulton2021}, the giant planet distribution peaks at $3.6^{+2.0}_{-1.8} \rm AU$ and afterwards decays as a power law $\propto a^\beta$ with $\beta=-0.86\pm0.41$, consistent with a log-uniform distribution ($\beta=-1$). The occurrence rate then decays for separations above $10\ \rm AU$, but suffers from completeness of the observed data. This present-day distribution may differ from the zero-age distribution. For simplicity, we choose a log-uniform distribution between $5\le a ({\rm AU}) \le 200$.

\textbf{Eccentricities}: For wide binaries, recent analysis by Gaia shows that the eccentricity distribution is consistent with a thermal distribution \cite{gaia21}. Very wide binaries could have slightly superthermal distributions \citep{hwang_e, ham_e}, but we keep the initial eccentricity distribution as thermal for simplicity. The planetary eccentricity is sampled from a Rayleigh distribution with $\sigma = 0.2$, as motivated by ref. \cite{Moorhead_2011_e=0.2_RayLeigh}, which studied the distribution of transit durations for Kepler planet candidates and their implications for orbital eccentricities. Similar initial Rayleigh eccentricity distribution indicates that the cold Jupiter population is consistent with high-e migration \cite{weldon2025}.

\textbf{Inclinations and other angles:} The mutual inclination between the planetary and the wide stellar orbit is uniform in cosine. In practice, each inclination $\iota_i$ is samples with respect to the plane of the galaxy. The mutual inclination is given by the cosine theorem 
\begin{equation}
    \cos \iota_{\rm mut} = \cos \iota_1 \cos \iota_2 + \sin \iota_1 \sin \iota_2 \cos (\Omega_1 - \Omega_2).
    \end{equation}
    The result is also uniform in cosine, as seen in Figure 7 in \cite{gp22}. This (cosine) uniform distribution is motivated by the recent analysis of \cite{christian2025} where all planets in binaries of separations above $700\ \rm AU$ have randomised angles. In fact, only the smaller planets ($R<6R_\oplus$) are aligned for binaries for separations below $700\ \rm AU$, and giant planets are misaligned for any separation. The other angles (arguments of pericentre $\omega_i$, argument of ascending node $\Omega_i$) are sampled uniformly.

The \textbf{stopping conditions} are similar to \cite{gp22}:
\begin{enumerate}
\item The system becomes dynamically unstable according to \cite{ma01}.
\item The planet reaches a pericentre small enough to be disrupted: $r_{\mathrm{peri}} < 2.7 a_{\rm Rohce}$, where $a_{\rm Roche}=R_p(M_\star/M_p)^{1/3}$, consistent with \cite{gui11}'s hydrodynamical simulations for planetary disruption.
\item The eccentricity of the planet, $e_1$, satisfies $e_1 \le 0.01$, and the final semi-major axis is 10\% smaller than the initial value.
\item The maximum simulation time of $13.5\ \rm Gyr$ or the main-sequence lifetime of the more massive star $t_{\rm ms}=10\ m^{-5/2}\ \rm Gyr$ has been reached, where $m=\max\{m_1,m_2\}/M_\odot$.
\end{enumerate}

\subsection*{Effect of flybys}
Here we quantify the effect of flybys on wide binaries. First, the widest binaries are prone to disruption. The typical half-life time for a wide binary is about $t_{1/2} \approx 10\ \rm Gyr (a/3\cdot 10^4\ \rm AU)^{-1}$ \cite{elbadri18}, consistent with the observed decrease of binaries with projected separations beyond $0.15\ \rm pc$ and in line with our upper limit.

Second, two body relaxation can lead to diffusion of the outer orbit to highly eccentric orbits with pericentre $q_2\ll a_2$. The diffusion rate depends on the ratio of the diffusion to the depletion. For slow diffusions (i.e. the diffusion timescale is longer than the outer period), the so-called empty loss cone regime is relevant, and the probability for a close encounter with pericentre $q_2$ is \cite{kaib14}

\begin{align*}
P(a_{2}|q_{2}) & =tn_{\star}a_{2}\sqrt{\frac{108\pi^{2}G\bar{m}q_{2}}{5}}=C_{12}a_{2}q_{2}^{1/2}\\
 & \approx0.1\left(\frac{q_{2}}{10\ {\rm AU}}\right)^{1/2}\left(\frac{a_{2}}{3\cdot10^{4}\ {\rm AU}}\right)^{1/2},
\end{align*}
Where we've used the typical time $t=13.5\ \rm Gyr$, the stellar number density in the Solar neighbourhood $n_\star=0.1\ \rm pc^{-3}$ and the mean mass of the flybys $\bar{m}=0.5M_\odot$. 

For very wide binaries the depletion is slow and the loss cone is empty. The critical separation between the two regimes is $a_c \propto (q_2/n_\star^2)^{1/7} \approx 4.2\cdot 10^4\ \rm AU (q_2/ 10\ \rm AU)^{1/7}$, which is not relevant for our problem.

Assuming the sampling of the inner and outer planetary orbit is independent
(which is mostly true for wide outer binaries), and assuming that
dynamical instability occur at $q_{2}=\beta a_{1}$ we get that the
probability for a binary at semi-major axis $a_{2}$ to experience dynamical
instability is
\begin{align*}
P(a_{2}) & =\intop_{ a_{1,{\rm min}}}^{ a_{1,{\rm max}}}P(a_{2}|q_2 = \beta a_{1})\frac{{\rm d}a_{1}}{a_{1}} \Bigg/ \intop_{ a_{1,{\rm min}}}^{a_{1,{\rm max}}}\frac{{\rm d}a_{1}}{a_{1}}\\
 & =\left(\ln\frac{a_{{\rm 1,max}}}{a_{1,{\rm min}}}\right)^{-1}\sqrt{2\beta}\left[a_{1,{\rm max}}^{1/2}-a_{1,{\rm min}}^{1/2}\right]C_{12}a_{2}=C_{2}a_{2}
\end{align*}
and finally the probability for a close encounter in our initial sample:
\begin{align*}
P_{{\rm enc}} & =\intop_{a_{2,{\rm min}}}^{a_{2,{\rm max}}}P(a_{2})\frac{{\rm d}a_{2}}{a_{2}}\Bigg/\intop_{a_{2,{\rm min}}}^{a_{2,{\rm max}}}\frac{{\rm d}a_{2}}{a_{2}}\\
 & =C_{2}\left(\ln\frac{a_{{\rm 2,max}}}{a_{2,{\rm min}}}\right)^{-1}\left[a_{2,{\rm max}}-a_{2,{\rm min}}\right]
\end{align*}
In deriving the latter equations we assume that the sampling is done
log-uniformly, which is indeed the case. Pugging in the values from
the simulation: $a_{1,{\rm min}}=5\ {\rm AU}$, $a_{1,{\rm max}}=200\ {\rm AU}$, and
$a_{2,{\rm min}}=300\ {\rm AU}$, $a_{2,{\rm max}}=3\cdot10^{4}\ {\rm AU}$, we
obtain $P_{{\rm enc}}\approx0.032\beta^{1/2}$. Even under the conservative assumption that $q_2 \approx 3a_1$ leads to instability, the overall probability is around $5.6\%$ of the systems, compared to a $20\%$ of dynamical instability achieved by galactic tides.

\subsection*{Exoplanet archive TAP search and data selection}

We obtain publicly available exoplanet data from the Exoplanet archive database: \url{https://exoplanetarchive.ipac.caltech.edu/TAP/sync}. We only retrieve exoplanets with masses $0.3\,M\sub{J}\le M\sub{p} \le 13\,M\sub{J}$ for which the host star's age and planetary radii are known, and stellar masses are in the range \mbox{$0.5\,M_\odot < M_\star \le 1.5\,M_\odot$}. We use the following ADQL query:

\begin{lstlisting}[float, language=ADQL]
SELECT DISTINCT pl_name, pl_orbsmax, pl_bmassj, pl_orbeccen, pl_orbper, st_mass, st_rad, st_age, st_ageerr1, st_ageerr2, pl_radj, pl_radjerr1, pl_radjerr2
FROM ps
WHERE pl_bmassj BETWEEN 0.3 AND 13
  AND st_mass BETWEEN 0.5 AND 1.5
  AND pl_orbsmax IS NOT NULL
  AND st_age IS NOT NULL
  AND st_ageerr1 IS NOT NULL
  AND pl_radj IS NOT NULL
  AND pl_radjerr1 IS NOT NULL
\end{lstlisting}

We follow by further slicing the period to $P<10\ \rm days$ and $e<0.1$, returning $319$ planets.

\subsection*{Secular Chaos}

 Even with single stars, secular chaos in the planetary systems can result in long term chaotic evolution and HJ formation due to overlapping secular resonances. Wu and Lithwick \cite{wu11} first demonstrated the numerical feasibility in producing HJs by running an example up to $0.4$ Gyr. Teyssandier et al. \cite{teyssandier19} used a secular model and accounted for dynamical and chaotic tides. Only a small fraction of the observed HJs can be reproduced in their sample. Moreover, the integrations were terminated by $t_{\rm end} = 2\ \rm Gyr$ due to numerical instabilities, with formation times peak at $\le 100$ Myr, and than stays flat in $\rm log(t)$. Teyssandier et al. argued that at most $10\%$ of additional HJs were missed by cutting the integration time 2 Gyr, which is incompatible with the recent evidence of later arrival \cite{hj_gyr, chen25}.

The only simulations that perform secular evolution up to the Hubble time are studied by Hamers et al. \cite{hamers17chaos}. The HJ formation time in their case has about $80\%$ of systems form at times $< 2$ Gyr (See inset of their Figure 9, reproduced for our Figure \ref{fig:4}). In our models, the $80\%$ threshold reached at $\sim 3.5$ and $\sim 7.5$ Gyr without and with galactic tides, respectively. Thus, wide-binaries even without galactic tides can deposit HJs at later times than secular chaos. We conclude that secular chaos can have significant contribution for HJ formation at times $0.1-1$, Gyr, but cannot be the source of later formations, incompatible with the recent claims that  most HJs were cool giant planets for more than one Gyr $\cite{old_hj26}$.

\subsection*{Giant Planet Mass-Radius Relation}

 Bashi et al. (2017) \cite{bashi_17_jup_mass_radius} found the mass-radius relationship for giant planets $R\propto M^{0.01\pm0.02}$. Recently updated, M\"{u}ller et al. (2024) \cite{halled_24_jup_mass_radius} found $R\propto M^{-0.06\pm0.07}$, both consistent with flat ratios. Already formed HJs are highly-irradiated by the host star, so their radii may be inflated \cite{komacek_inf}. Young Jupiters, in contrast, may still be thermally contracting. This will affect their coupled tidal evolution as it is a strong function of the planetary radius \cite{roz22, glanz22}. However, since the typical HJ-formation timescales are of the order of Gyr, we fix the radius at $1 R_J$ for simplicity and do not attempt to adjust for the post HJ-formation inflation, which is beyond the scope of this work. We finish this section by comparing the sample of HJ with the cold Jupiter sample that consists of 95 planets with known stellar ages and planetary radii. Cold Jupiters are usually discovered via radial velocities where radii are unknown, so this sample is far from complete.

\begin{figure*}
    \centering
    \includegraphics[width=0.99\linewidth]{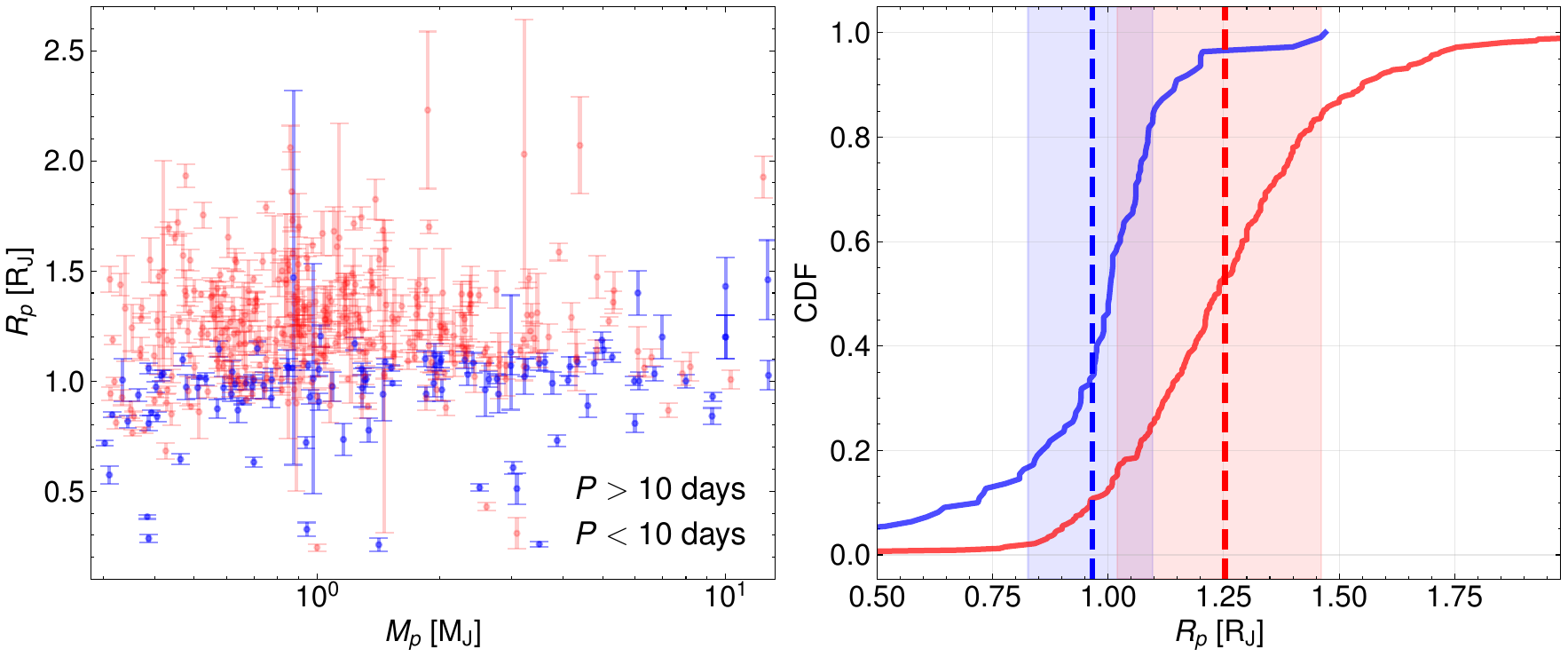}
    \caption{Mass-radius relations of HJs. Left: Mass-Radius distribution for HJ ($P<10\ \rm days$, red) and cold/warm Jupiters ($P>10\ \rm days$, blue). Right: CDFs of the data on the left panel. Dashed lines are the mean values and the shaded regions represent the $1-\sigma$ confidence levels.}
    \label{fig:ex2}
\end{figure*}

The left panel of Extended Data Figure \ref{fig:ex2} shows the mass-radius relationship for HJs ($P<10\ \rm days$ and $e<0.1$, red) and Cold/warm Jupiters ($P>10\ \rm days$, blue). It is worth noting that Cold Jupiters have radii close to $\sim 1R_J$ and, as expected, they are independent of the mass \cite{bashi_17_jup_mass_radius, halled_24_jup_mass_radius}. On the contrary, the HJ sample has a much larger mean radius. This is also evident from the right-hand panel, which shows the CDF for both populations. In this panel, dashed lines are mean values and the shaded regions represent the $1-\sigma$ confidence levels. The Cold and hotJupiter populations have $R_{\rm CJ}=0.96^{+0.14}_{-0.16}\,R_{\rm J}$ and $R_{\rm HJ}=1.25^{+0.19}_{-0.23}\,R_{\rm J}$, respectively. The stellar ages and uncertainties of the 319 HJ systems were previously analysed and plotted in Figure 4.

\subsection*{Comparing models for late arrived hotJupiters}

\begin{figure*}
    \centering
    \includegraphics[width=0.99\linewidth]{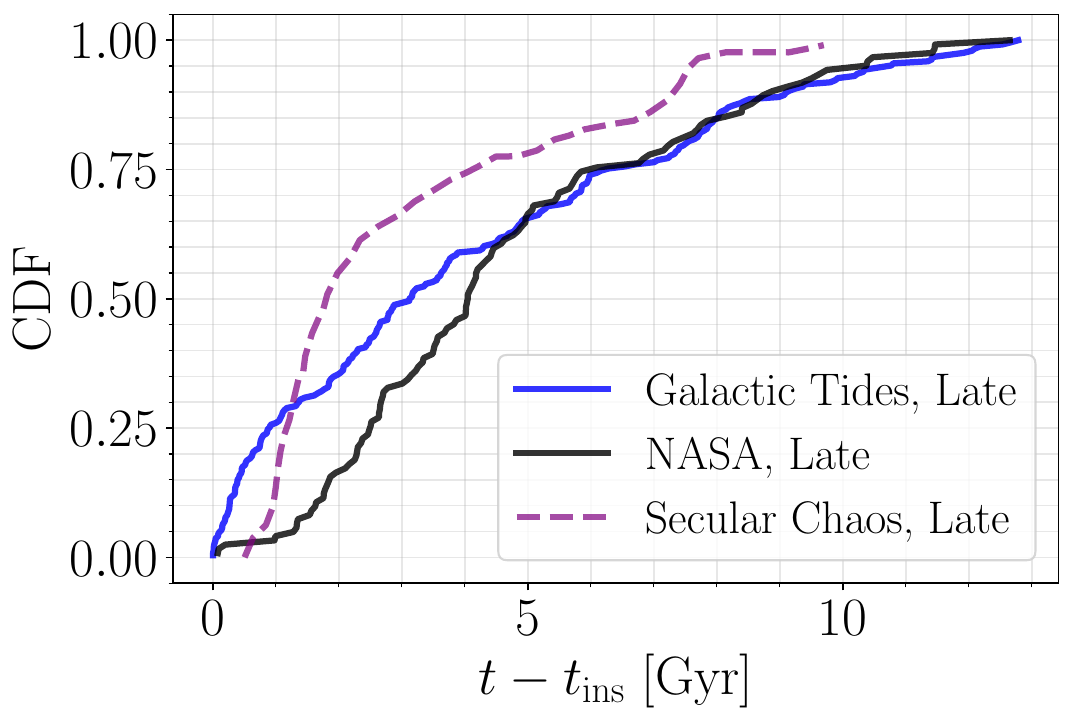}
    \caption{Delay time distributions for late arrived HJ where the inspiral time is subtracted. Black line indicates the HJ sample with known median stellar ages from the NASA exoplanet archive database. Dashed purple line shows the result of the secular chaos model \cite{hamers17hj} Solid blue line indicates our model with GT.}
    \label{fig:ex3}
\end{figure*}

While this work was under review, Chen et al., (2025) \cite{chen25} provided further evidence that $38^{+16}_{-14}\%$ has arrived late. Their definition of late arrival is that the inspiral time $t_{\rm ins}$ is shorter than the stellar age at the 2-$\sigma$ level. For a HJ the inspiral time is given by 
\begin{equation}
    t_{\rm ins} = \frac{2}{13}\frac{2Q'_\star}{9}\frac{M_\star}{M_{\rm J}}\left( \frac{a}{R_\star} \right)^5 \frac{P}{2\pi},
\end{equation}
where $M_\star$ and $M_{\rm J}$ are the stellar and Jovian masses, $R_\star$ is the stellar radius, $a$ is the semi-major axis and $P$ is the orbital period. $Q'_\star$ is the tidal quality factor and has large uncertainty. Reported values in the literature span five orders of magnitude from $10^4$ to $10^9$. We take $Q'_\star=10^6$ to be conservative and compatible with the 1-$\sigma$ higher end with the empirically derived value of $\log Q'_\star=5.7^{+0.4}_{-0.3}$ (see Chen et al., 2025 \cite{chen25} for details). After applying this cut we're left with $122$ systems. For our model, there is no uncertainty on the arrival time and our cut is simply $t - t_{\rm ins} > 0$.

\begin{figure*}
    \centering
    \includegraphics[width=0.99\linewidth]{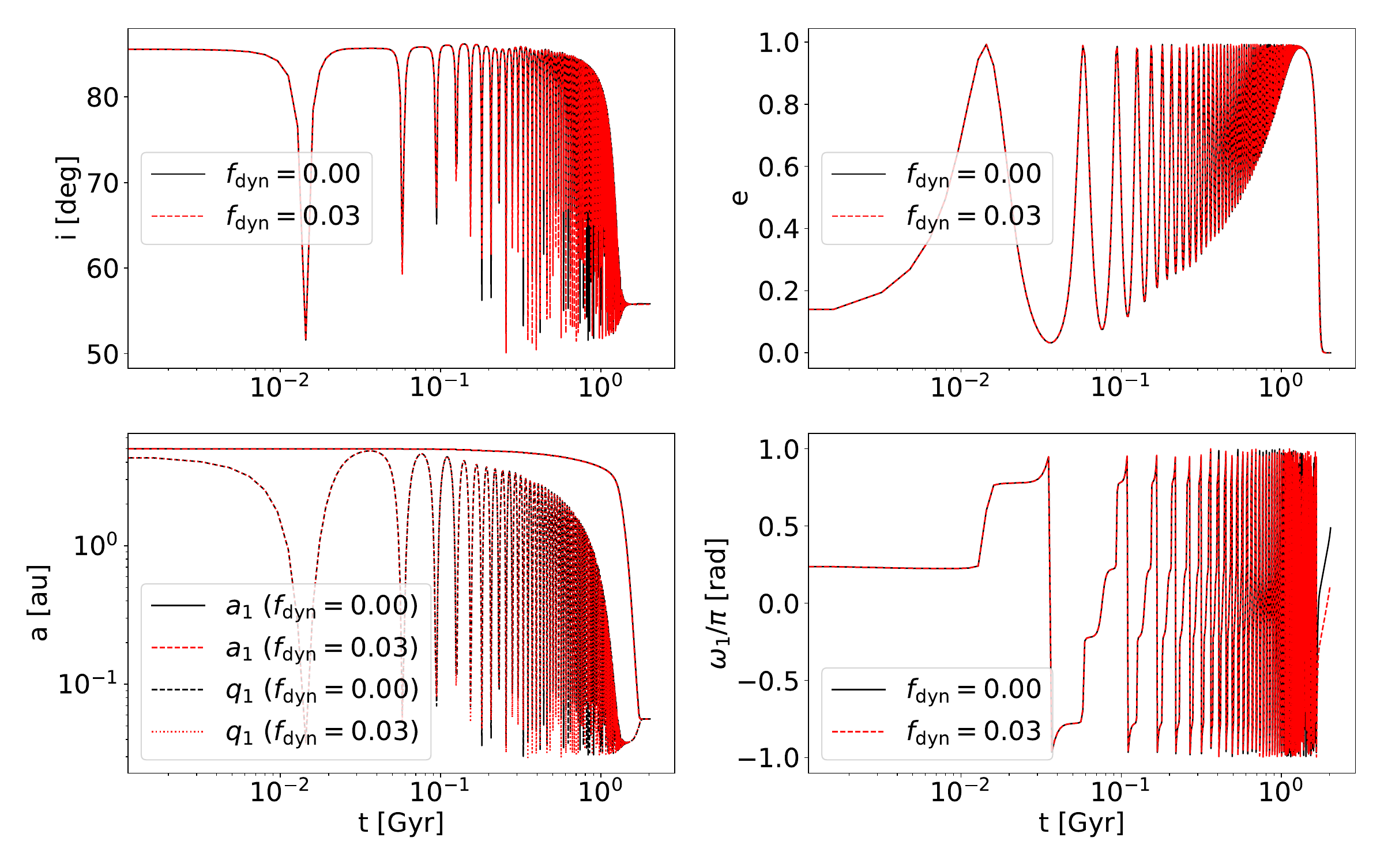}
    \caption{Comparison of orbital evolution for a HD 80606b-like system with different dynamical friction parameters, \( f_{\mathrm{dyn}} = 0.00 \) (black solid lines) and \( f_{\mathrm{dyn}} = 0.03 \) (red dashed lines). The plots show inclination (\(i\)), eccentricity (\(e\)), semi-major axis (\(a_1\) and pericenter \(q_1\)), and argument of periapsis (\(\omega_1/\pi\)) as functions of time. The close agreement indicates that moderate changes in \( f_{\mathrm{dyn}} \) do not significantly affect the orbital evolution over the timescales of our simulations.}
    \label{fig:ex4}
\end{figure*}

\begin{figure*}
    \centering
    \includegraphics[width=0.99\linewidth]{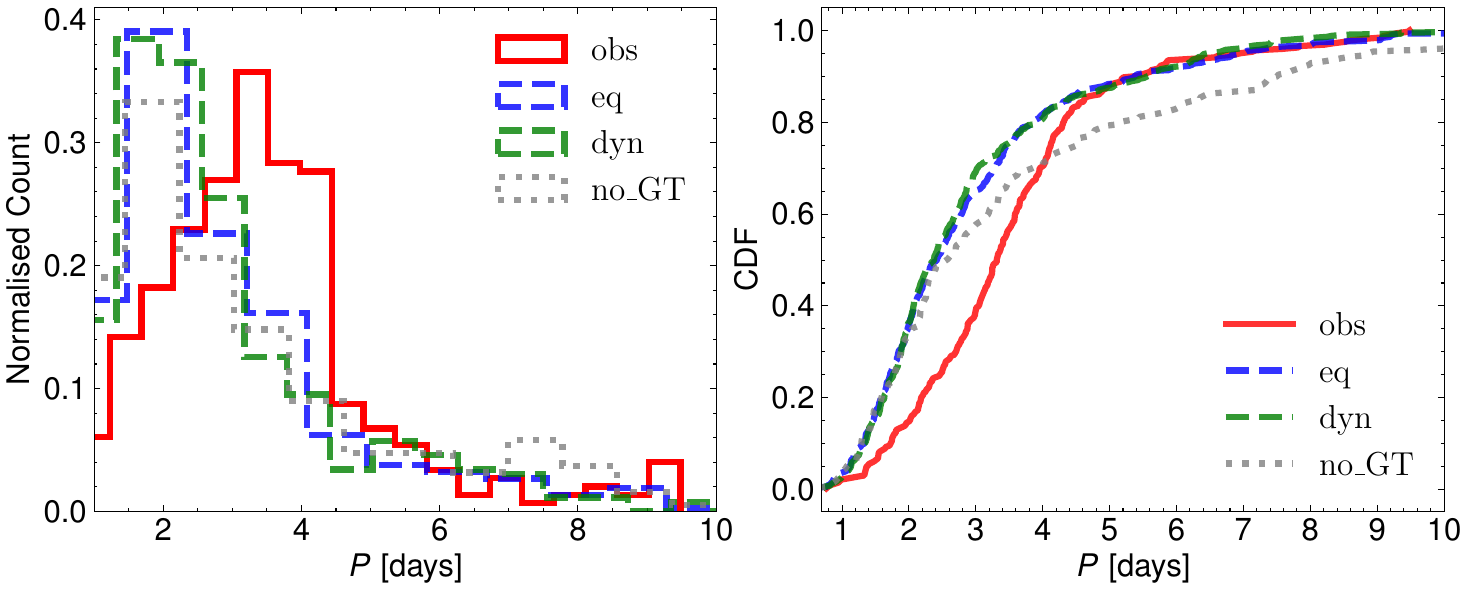}
    \caption{Simulated and observed period distributions. Left: normalised histogram of the period distribution of the observed sample (solid red lines), the simulated models with equilibrium (dashed blue), dynamical (dashed green) and the equilibrium tide without GT (dotted lines). Right: The same data but for the cumulative distribution function (CDF).}
    \label{fig:ex5}
\end{figure*}

Extended Data Figure \ref{fig:ex3} shows the CDF of the arrival times of the later arrived HJs from the NASA database (black soid line), our GT model (solid blue) and the secular chaos model (dashed purple). The data for secular chaos is reproduced from the Extended Data Figure \ref{fig:ex2} in Chen et al., (2025) \cite{chen25} using \href{https://automeris.io/docs/digitize/#input-format-for-calibration-values}{WebPlotDigitizer}. We see that our GT model is superior to the secular chaos model for systems that arrived later than 1 Gyr after accounting for tidal inspiral  ($t-t_{\rm ins}>1 \ \rm Gyr$).

These results should be taken with caution. Our observed sample is from the public NASA exoplanet database, while Chen et al., (2025) \cite{chen25}'s sample is from kinematic catalogues. Also the details of the tidal model are highly uncertain (see below), and $Q'_\star$ can vary significantly between individual systems depending on the stellar structure which can lead to a large variation in the estimation of $t_{\rm ins}$. Nevertheless, either in the raw data or only for late arrivals, the NASA sample tends to favour our galactic tide model over the secular chaos model. In reality, the origins of old HJ will be some combination of both (and possibly other) channels. 

\subsection*{Dynamical tides in eccentric binaries}

We have used a simplified model for equilibrium tides based on a fixed value of the apsidal motion constants and the viscous times \citep{hut81}. The apsidal motion constants $k_p=0.37$ and $k_\star = 0.028$ were taken from ref. \citep{anderson2016}. For the viscous time we use $t_{\nu \star}=50\ \rm yr$ and $t_{ \nu,p}=0.01\ \rm yr$, following ref. \cite{weldon2025} which follows ref. \cite{petrovich2015}.

We also implement a model for dynamical tides in highly eccentric binaries following refs. \cite{mk18, gp22}. In particular, we follow Grishin and Perets (2022) \cite{gp22} and keep both tidal prescriptions. There will be an effective eccentricity $e_{\rm eff}$ where the equilibrium tide will dominate below $e<e_{\rm eff}$ \cite[see Sec. 2.4.3 in][]{gp22}. This is contrary to a switch at $e=0.8$ implemented in \cite{mk18} that makes the tidal prescription discontinuous. In particular, the energy loss during the pericentre passage due to dynamical tides raised on body 2 (the planet) by body 1 (the star) is
\begin{equation}
    \Delta E_{1\to2} = f_{\rm dyn} \frac{m_1+m_2}{m_1}\frac{Gm_2^2}{R_1}\left( \frac{R_1}{a_1(1-e_1)} \right)^9,
\end{equation}
with a similar expression for $\Delta E_{2 \to 1}$ (keeping the pericentre term $a_1(1-e_1)$ unchanged). The efficiency of the dynamical tide is encapsulated in $f_{\rm dyn}$. We ran an additional set of $5000$ simulations with $f_{\rm dyn}=0.03$ and did not find significant changes for the population analysis. We obtained $426$ HJs, $450$ disruptions, and $964$ unstable giant planets, which is consistent with the run without including dynamical tides. These results are also supported by previous findings, showing that the efficiency of tidal migration is essentially independent on the tidal model selected \citep{kaib14,petrovich2015}. For a quantitative test case, we have reproduced the evolution of HD 80606b with and without dynamical tides, using the same initial conditions and tidal parameters as in ref. \citep{Wu&Murray}. We did not find any significant difference in the evolution as shown in Extended Data Figure \ref{fig:ex4}.

Nevertheless, more detailed modelling of dynamical and chaotic tides will be necessary. The dominant $f$-mode energy can random-walk to large values and could be efficient in extracting orbital energy from very eccentric objects near their Roche limit \citep[e.g.,][]{mardling95a, mardling95b,vl17}. The resulting model could prevent some HJs from undergoing tidal disruption. While \cite{wu18} argues that most proto-HJs could avoid tidal disruption and efficiently decouple from the wider binary, the population study by \cite{vl19} found a modest increase in the HJ formation rate, explaining at most $20-30\%$ of HJs. If indeed tidal dissipation is efficient, the combined chaotic tidal evolution and the inclusion of GT in wide binaries could explain most HJs. This may also shed light on the formation of wide-binary planets such as TOI-5688 A b \citep{Reji2025}, part of the growing number of giant exoplanets orbiting M-dwarfs \citep[GEMS;][]{Glusman2026}. However, efficient tidal evolution may also skew the delay-time distribution so that HJs will form after a few ZLK cycles. This would pose a challenge to the late-forming HJs such as NGTS-10b \citep{ngts-10}, or TOI-2019b which is found around a star with $t_{\rm age} = 1.77^{+0.88}_{-0.68}$ Gyr and is expected to disrupt in a few Myr \citep{Alvarado-Montes2025}, as opposed to GEMS planets whose final tidal evolution can last for several Gyr \cite{Alvarado-Montes2022}.

\subsection*{Period Distribution} 

We compare the observed period distribution of HJs with the results from our simulations. As seen in the histogram on the left panel of Extended Data Figure \ref{fig:ex5}, the simulated HJs (for both equilibrium and dynamical tide modes) have shorter periods than the observed ones. The right panel of Extended Data Figure 5 shows the CDF of the same data. Although the simulated data has indeed more short period orbits, the HJ population with orbits longer than $\sim 4.5\ \rm days$ is compatible with the simulated data. The model without GT (dotted grey) has similar features to the other simulated models, but can extend to slightly longer periods.

The discrepancy may be explained by several effects: As discussed in the initial condition for the population study, the observed masses are different from the sampled masses and suffer from observational biases. The masses affect the period distributions via the strength of the tidal dissipation and setting the Roche limit for a disruption event in lieu of a HJ formation. Additionally, radius inflation \cite{roz22, glanz22, lu25} may save some of the disrupting planets and shift the period distribution to larger values which are more consistent with the observations. Finally, angular momentum return due to planetary mass loss \cite{weldon26} can increase the survival fraction of HJ by a factor $2-3$, potentially explaining the majority of HJs.

Additional effects we have not included are dynamical tides in the form of inertial waves in convective envelopes and internal gravity waves in stellar radiative regions, which operate even for circular and spin-synchronised HJ systems. These effects induce a fast tidal disruption for some of the simulated HJs, as observed in TOI 2019b and NGTS-10 \cite{Alvarado-Montes2025, ngts-10}, which will eliminate HJs on short period orbits that formed too early from the observed sample. A self-consistent analysis that consists of eccentric tidal migration, subsequent tidal evolution and hierarchical Bayesian modelling that estimates the observability of the simulated data, its implications, and comparison to the observed stellar age and orbital distribution is planned for future work (Alvarado-Montes, Grishin et al., 2026, in prep.).

\subsection*{Data availability}
The simulated data is available at (\url{https://github.com/eugeneg88/SecuLab}). The observational data is also public and avaiable in the NASA Exoplanet Archive database (\url{https://exoplanetarchive.ipac.caltech.edu/TAP/sync}). We've provided the ADQL query to get updated data in the methods. 

\subsection*{Code availability}
The secular code is public and is available at (\url{https://github.com/eugeneg88/SecuLab}). 

\section*{Acknowledgements}
We thank Sharan Banagiri, Alexey Bobrick, Ryosuke Hirai and Hagai B. Perets for useful discussions. EG acknowledges support from the ARC Discovery Early Career Research Award (DECRA) DE260101802. JAAM acknowledges support from the Macquarie University Research Fellowship (MQRF) scheme.

\section*{Author contributions}
EG conceived the idea, wrote the secular code, ran the population study and interpreted the results. JW ran the individual cases in Figure \ref{fig:2}, contributed to coding the initial conditions for the population study, plotting, and interpreting the results. JAAM contributed to discussing potential observational implications of the population study. EG, JW, and JAAM contributed to reviewing and editing the manuscript.

\section*{Competing Interests}
The authors declare no conflict of interest.

\bibliographystyle{naturemag}

\end{document}